\documentclass[review]{elsarticle}

\usepackage{lineno,hyperref}
\journal{Journal of \LaTeX\ Templates}

\begin{document}

\begin{frontmatter}

%\title{Variation in energy stored and dissipated in hysteretic system in applied field with relative phase of sinusoidal components of field}
\title{Variation in energy stored and dissipated in type-II superconductor in applied ac magnetic field with relative phase of two sinusoidal components of the field}


\author{Zden\v{e}k Jan\accent23 u}
\author{Tymofiy Chagovets}
\address{Institute of Physics of the Czech Academy of Sciences, Na Slovance 2, 182 21 Prague 8, Czech Republic}




\begin{abstract}
We show that both the energy stored and dissipated by a system with hysteretic nonlinearity in an applied field varies with the relative phase of the sinusoidal components of the field, even if the magnitude of these components, and thus an effective value of the field, are kept constant. The explored system is a type-II superconductor in the critical state subjected to a time varying applied magnetic field. Complete analytical expressions for hysteresis loops, determined from basic physical phenomena, are known for this system. A theoretically predicted variation in the energy is in good agreement with our experimental measurements.
\end{abstract}

%\begin{keyword}
%\texttt{elsarticle.cls}\sep \LaTeX\sep Elsevier \sep template
%\MSC[2010] 00-01\sep  99-00
%\end{keyword}

%\pacs{75.60.-d, 74.25.Uv, 74.25.Wx}
%Nonlinear dynamics, 05.45.-a
%Hysteresis in magnetism, 75.60.-d
%Magnetization curves, hysteresis, Barkhausen and related effects 75.60.Ej (for hysteresis in ferroelectricity, see 77.80.Dj)
%Granular systems, classical mechanics of, 45.70.-n
%Sand piles, phase transitions in, 64.60.av
%Vortex phases (superconductivity), 74.25.Uv
%Vortex pinning (superconductivity), 74.25.Wx
%Magnetization of superconductors, 74.25.Ha
%Critical fields (superconductivity), 74.25.Op
%Pattern formation in complex systems, 89.75.Kd
%\keywords{hysteresis, nonlinearity, energy, dissipation, superconductivity}

\end{frontmatter}

%\linenumbers

\section{Introduction}

There are many systems showing hysteretic nonlinearity - rate independent memory effects. For ferromagnetic, ferroelectric or elasto-plastic materials etc. models of hysteresis such as the empirical Preisach or Jiles-Atherton model are employed \cite{Bertotti2006,Visintin2014,Preisach1935,Jiles1984}. However, complete analytic expressions for hysteresis loops, worked-out from basic physical phenomena, are rather rare. Such a case is a magnetic moment of an induced shielding supercurrent in a type-II superconductor in the critical state subjected to an applied time-varying magnetic field \cite{Campbell2001,Hughes1974}. The magnetization hysteresis loops are independent of field rate at low enough rates and when the field is cycled so that only major-loops are formed, the loops depend on the peak value of the field but not on the field waveform.

Hysteretic nonlinearity produces non-linear effects such as harmonics of the fundamental-frequency when a pure sinusoidal field is applied and frequency mixing when two sinusoidal field components are applied. When, for example the components at the fundamental-frequency and third-harmonic-frequency with fixed magnitudes are applied concurrently, the loops depend on a relative phase $\varphi$ of these components, even if the effective value of the total field is constant. This is a consequence of a variation in the peak value of the field with $\varphi$. Hence, a mean stored energy of the magnetic moment in the applied field and energy dissipated per fundamental field cycle depend on $\varphi$, too. This is a fundamental difference to linear systems, where a superposition principle holds and the relative phase of the field components plays no role.

Type-II superconductors remain in the mixed state even in high magnetic fields owing to the creation of quantized vortices (vortex lines) \cite{Campbell2001,Hughes1974}. Each vortex carries a single quantum $h/2e$ of magnetic flux. The vortices or anti-vortices, according to the direction of the applied field, appear at an edge of a sample and propagate towards the sample's interior to homogenize its density. The moving vortex produces a time-varying magnetic field in the sample and thus dissipates electromagnetic energy. %and produces Joule heat.
Artificial defects of sub-micron scale are intentionally produced in the superconductor within its production process to pin the vortices in the material and thus prevent energy losses due to the relaxation of the magnetic flux density profiles \cite{Campbell2001,Hughes1974,Selvamanickam2015}. The strength of the pinning force determines the critical depinning current density $j_c$, the maximum dc current density that can flow in the superconductor without energy dissipation. When transport or induced electric current $j$ flows in a sample, the Lorentz force acts to unpin the vortices and gives rise to vortex diffusion or flow. The critical state occurs when the Lorentz force and pinning force are balanced \cite{Campbell2001,Hughes1974}. The current density and vortex density profiles rearrange in a such way that the maximum current density is $|j| \le j_c$ in the whole sample and the flux density profiles are with the gradient $\nabla B = \pm \mu_0 j_c$. Since the screening supercurrent is persistent and a distribution of the current density and profiles of the flux density are quasi-static, work done by the Lorentz force acting on the vortices against the pinning force is field rate independent, so the problem is analogous to a system with dry friction-like pinning.

On the basis of the critical state theory, distribution of the current density in the sample and generated magnetic moment $m$ may be expressed as a function of the applied field $B_a$ and $j_c$. Analytical solutions for $m(B_a)$ are known for restricted shapes of the samples and configuration of the samples in the applied field, namely when spatial symmetry reduces the problem from 3D to 2D. Infinite slabs and cylinders in a parallel field fall into first group \cite{Goldfarb1991}. Flat strips with a length $l$, width $2a \ll l$ and thickness $d \ll a$, and circular disks (pucks) with a radius $R$ and thickness $d \ll R$ oriented perpendicularly to the applied field fall into the second group \cite{Brandt1993b,Clem1994}. When the magnetic field $B_a$ is applied perpendicularly to the flat surface of the strips or disks, the screening currents generate a magnetic moment with the component $m$ parallel to $B_a$ only.

\section{Modelling}

The initial magnetization curve for the strip is given by

\begin{equation}\label{m(B)StripCurve}
m_{i}(B_a) = -(\pi a^2 l/\mu_0) B_d \tanh(B_a/B_d),
\end{equation}

\noindent where $B_d = \mu_0 j_c d/ \pi$ is the characteristic field \cite{Brandt1993b}. The initial magnetization curve determines the full magnetization hysteresis loop,

\begin{equation}\label{m(B)StripLoop}
m_{\downarrow\uparrow} (B_a) = \pm m_i \left( \frac{B_p}{B_d} \right) \mp 2 m_i \left( \frac{B_p \mp B_a}{2 B_d} \right),
\end{equation}

\noindent where a branch $m_\downarrow$ is for the field $B_a$ monotonously non-increasing from $B_p$ to $-B_p$ and branch $m_\uparrow$ is for the field $B_a$ monotonously non-decreasing from $-B_p$ to $B_p$. The shape of the magnetization loops depends only on the ratio between the peak value $B_p$ of the applied field and the characteristic field $B_d$. Owing to the large aspect ratio, $a/d$ or $R/d$, of the flat samples in the transverse field the magnetization loops for the disks and strips, normalized to the magnetic moment of the sample with perfect screening% (and saturation value)
, differ by less than 1.2\%. Numerical modelling shows the same similarity of the loops for flat rectangles and hence these may be well approximated by analytical expressions for the disks or strips \cite{Brandt1997}.

Since both applied field $B_a(t)$ and induced magnetic moment $m(t)$ are periodic functions of time, we use a discrete fast Fourier transform to analyse these waveforms. The complex amplitude $\mathcal{B}(f_1)$ at the fundamental frequency $f_1$ is a Fourier coefficient of the applied pure sinusoidal field $B_a(t)= B_1 \sin ( 2 \pi f_1 t)$. Hysteretic nonlinearity of the magnetization loops $m(B_a)$ generates the components of the magnetic moment at harmonics of the fundamental frequency. We apply the notation $\mathcal{M}_{n}(f_1) \equiv \mathcal{M}(nf_1)$ for the $n$th harmonic component of the magnetic moment. As the symmetry of the magnetization loops is odd in the applied field with zero dc component, only odd harmonics appear in the $\mathcal{M}_n(f_1)$.

\section{Experimental}

Experimental data were produced by measurements of the 4 mm long cut of a 4 mm wide sample of a second-generation high-temperature superconducting wire \cite{Superpower}. The magnetization loops were measured in a continuous reading SQUID magnetometer \cite{Janu2014,Janu2015}.

\begin{figure}
\includegraphics[scale=0.45]{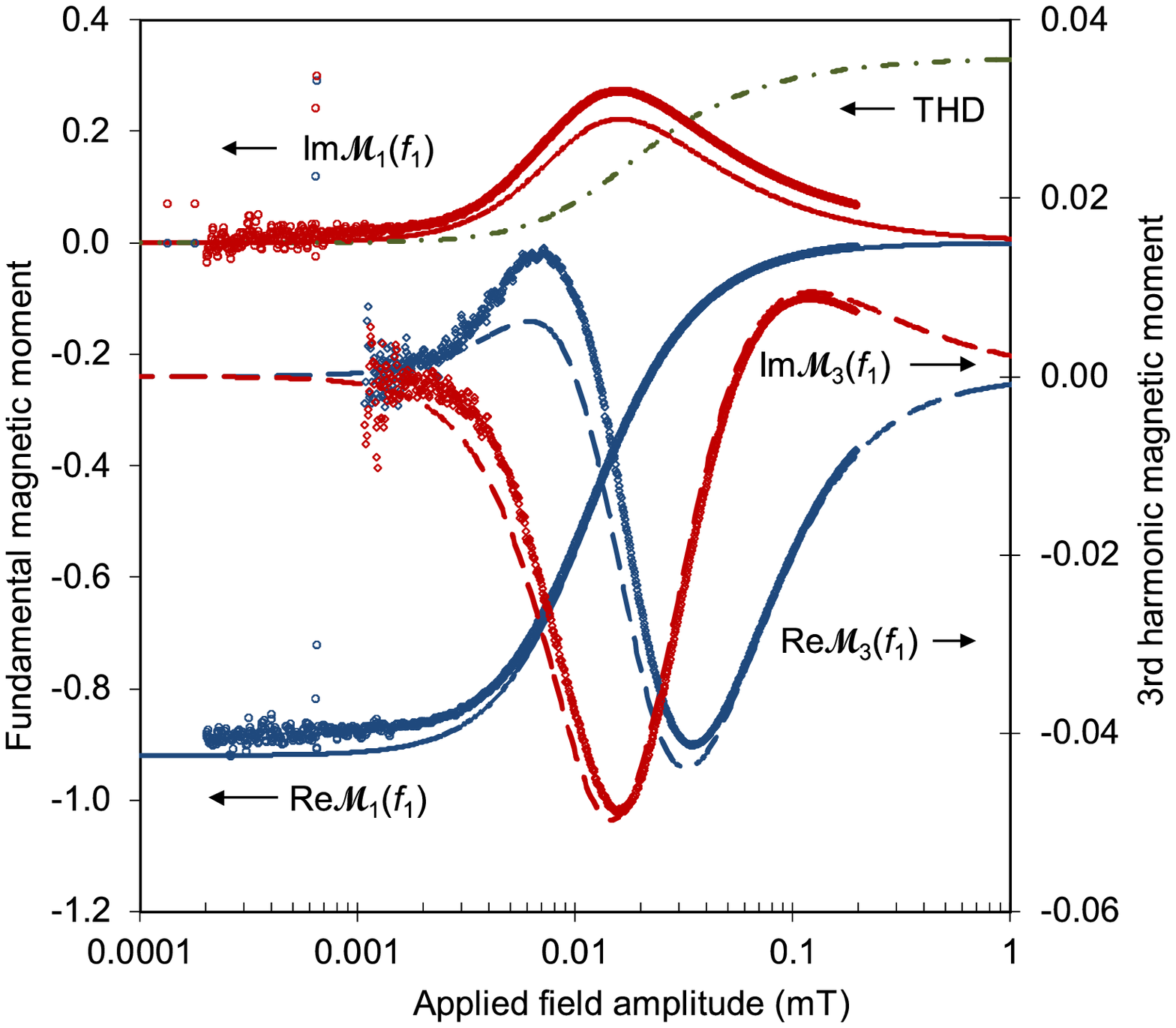}% Here is how to import EPS art
\caption{\label{m(B)} Real and imaginary parts of the fundamental-frequency and third-harmonic-frequency complex amplitudes of the magnetic moment, normalized to the magnetic moment of the sample with perfect screening, as a function of the applied field amplitude. Symbols are for experimental data. Curves represent the amplitudes calculated on the basis of the model. The dash-dot curve represents nonlinearity expressed by a "third harmonic distortion", see text% of the magnetic moment%$|\mathcal{M}_3(f_1)|/|\mathcal{M}_1(f_1)|$ of the induced magnetic moment
.}
\end{figure}

Fig. \ref{m(B)} shows the theoretical and experimental amplitudes of the magnetic moment at the fundamental-frequency and third-harmonic frequency as a function of the amplitude of the applied pure sinusoidal field. The amplitudes are normalized to the magnetic moment of the sample with perfect screening (Meissner state) \cite{Brandt1997}. The theoretical amplitudes are the Fourier coefficients of the magnetization loops given by Eq. \ref{m(B)StripLoop} which were calculated for discrete values of $B_p/B_d$, where $B_p = B_1$. The experimental amplitudes are the Fourier coefficients of the induced magnetic moment
%measured in a pure sinusoidal field with an amplitude $B_1$ swept exponentially at a frequency of $f_1 = 1.5625$ Hz and constant temperature 89.945 K
measured in a pure sinusoidal field with frequency $f_1= 1.5625$ Hz, amplitude $B_1$ swept exponentially and constant temperature 89.945 K, i.e. at constant $j_c$ \cite{Janu2014}. Distinct behavior of the amplitude at the 3rd harmonic frequency confirms the critical state with $j_c > 0$ in the sample. The theoretical pairs  $(cB_p/B_d,\mathcal{M}_n)$ were matched to the experimental pairs $(B_p,\mathcal{M}_n)$ by using a single parameter $c$ that offers contact-less determination of the critical depinning current density $j_c = c \pi/\mu_0 d$ \cite{Janu2014}.

%\begin{figure}
%\includegraphics[scale=0.45]{Fig1}% Here is how to import EPS art
%\caption{\label{m(B)} Real and imaginary parts of the fundamental-frequency and third-harmonic-frequency complex amplitudes of the magnetic moment, normalized to the magnetic moment of the sample with perfect screening, as a function of the applied field amplitude. Symbols are for experimental data. Curves represent the amplitudes calculated on the basis of the model. The dash-dot curve represents nonlinearity expressed by a "third harmonic distortion", see text% of the magnetic moment%$|\mathcal{M}_3(f_1)|/|\mathcal{M}_1(f_1)|$ of the induced magnetic moment.}
%\end{figure}

%\begin{figure}
%\includegraphics[scale=0.45]{Fig1a}% Here is how to import EPS art
%\caption{\label{m(B)} Real and imaginary parts of the fundamental-frequency and third-harmonic-frequency complex amplitudes of the magnetic moment, normalized to the magnetic moment of the sample with perfect screening, as a function of the applied field amplitude. Symbols are for experimental data. Curves represent the amplitudes calculated on the basis of the model. The dash-dot curve represents nonlinearity expressed by a "third harmonic distortion", see text% of the magnetic moment%$|\mathcal{M}_3(f_1)|/|\mathcal{M}_1(f_1)|$ of the induced magnetic moment
%.}
%\end{figure}

We express nonlinearity by the "third-harmonic-distortion" $|\mathcal{M}_3(f_1)|/|\mathcal{M}_1(f_1)|$. Fig. \ref{m(B)} shows that in the weak field, $B_p \ll B_d$, hysteretic nonlinearity plays a secondary role. Nonlinearity increases %rapidly
in an environment of a peak of the imaginary %(out-of-phase)
part and drop in the real part of the normalized amplitude of the magnetic moment at fundamental frequency which occur for $B_p / B_d$ of the order of unity, when the critical-state profile penetrates just to the center of the sample \cite{Brandt1993b,Clem1994}. For $B_p \gg B_d$, nonlinearity saturates while the amplitudes of the magnetic moment drop to zero.

In a steady state, the mean stored energy of the induced magnetic moment in the applied field is $E'(f_1) = -\mathrm{Re}\left[\mathcal{B}^*(f_1)\mathcal{M}_1(f_1)\right]/2$, where the asterisk denotes the complex conjugate. The energy dissipated per cycle of the applied field is $E''(f_1) = \mathrm{Im}\left[\mathcal{B}^*(f_1)\mathcal{M}_1(f_1)\right]/2 = W(f_1)/2\pi$, where $W(f_1)$ is the area of the magnetization hysteresis loop. Since hysteresis is a rate independent effect, both mean stored energy and energy dissipated per field cycle are frequency independent. In spite of it, we use this notation for reasons that will become clear in following text.

The applied field is composed of two sinusoidal fields at the fundamental-frequency and third harmonic frequency,

\begin{equation}\label{Ba(t)}
B_a \left( t \right) = B_1 \sin{ \left( 2\pi f_1 t \right)} + B_2 \sin{\left(2 \pi 3 f_1 t + \varphi\right)},
\end{equation}

\noindent whose relative phase is $\varphi$. The Fourier transform of this field has the complex components $\mathcal{B}(f_1)$ and $\mathcal{B}(3f_1)$. The effective value of the total field is $B_{rms}=2^{-1/2}[|\mathcal{B}(f_1)|^2+|\mathcal{B}(3f_1)|^2]^{1/2}$. While the magnitudes of $\mathcal{B}(f_1)$ and $\mathcal{B}(3f_1)$ are kept constant, the field waveform and peak value of the field vary with the relative phase $\varphi$. Since the expression for the hysteresis loops, Eq. \ref{m(B)StripLoop}, holds on the assumption that the applied field $B_a$ is monotonously non-decreasing from $-B_p$ to $B_p$ and monotonously non-increasing from $B_p$ to $-B_p$, we have to put a restriction $|\mathcal{B}(3f_1)|/|\mathcal{B}(f_1)| \leq 1/9$ on the magnitudes to exclude the formation of the minor magnetization loops.

Now, because of hysteretic nonlinearity, the amplitudes $\mathcal{M}_n(f_1)$ and $\mathcal{M}_n(3f_1)$ of the magnetic moment are dependent on both sinusoidal components $\mathcal{B}(f_1)$ and $\mathcal{B}(3f_1)$ of the field and they vary with a relative phase $\varphi$ of these components while its magnitudes and thus the effective value of the field are kept constant. The energy of the magnetic moment in the applied field has the components $E(f_1) = - \mathcal{B}^*(f_1)\mathcal{M}_1(f_1)/2$ and $E(3f_1) = - \mathcal{B}^*(3f_1)\mathcal{M}_1(3f_1)/2$ which depend on $\varphi$, too. The total mean stored energy is $E' = E'(f_1) + E'(3f_1)$ and the total energy dissipated per cycle of the field $\mathcal{B}(f_1)$ is $E'' = E''(f_1) + 3E''(3f_1)$ because three cycles of the field component $\mathcal{B}(3f_1)$ occur during one cycle of the field component $\mathcal{B}(f_1)$.

\begin{figure}
\includegraphics[scale=0.45]{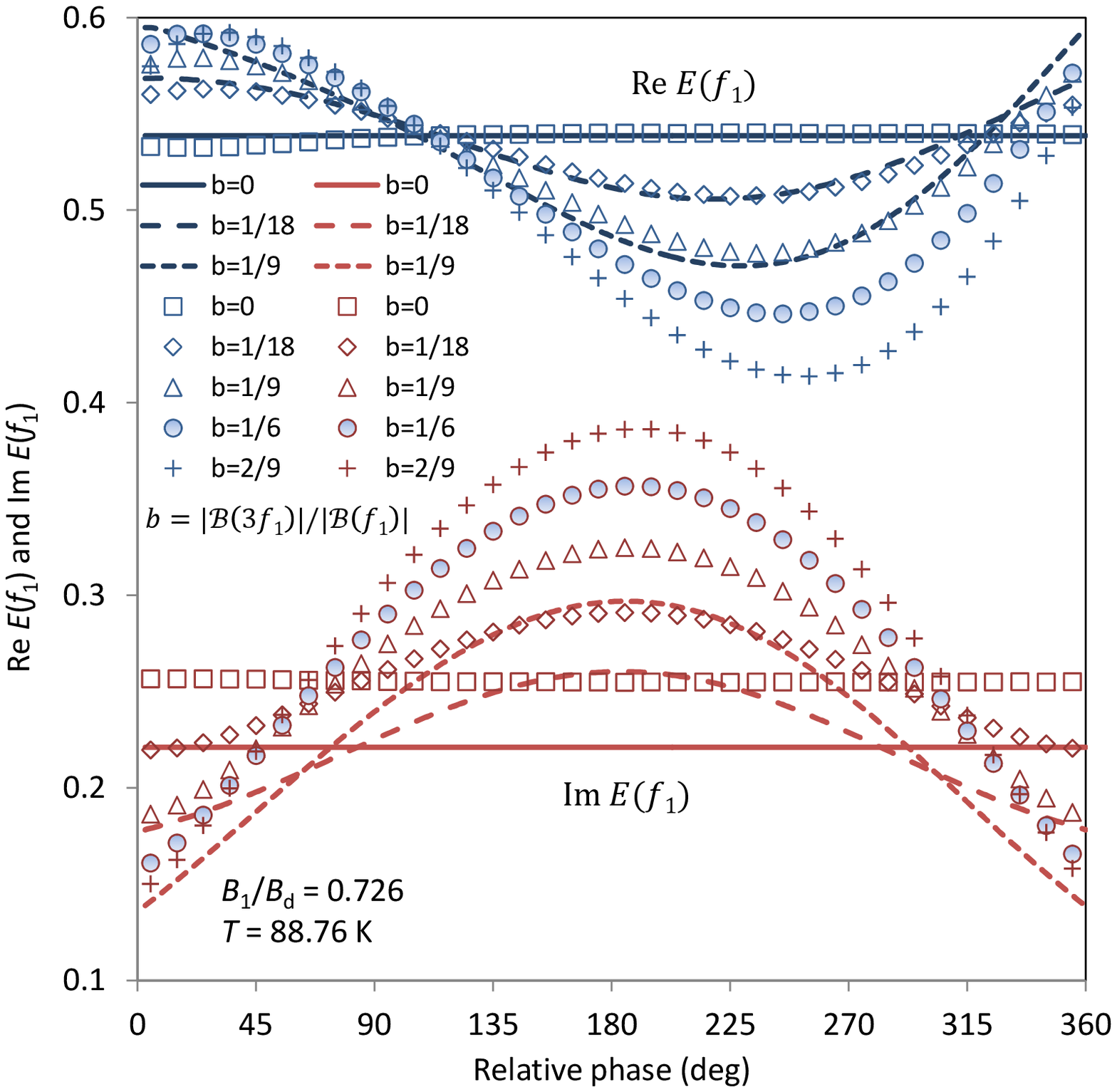}% Here is how to import EPS art
\caption{\label{E1(phi)} The energy of the induced magnetic moment in the applied field at fundamental-frequency vs. relative phase $\varphi$ of the field components $\mathcal{B}(f_1)$ at fundamental frequency and $\mathcal{B}(3f_1)$ at third harmonic frequency. The symbols are for the experimental data. The curves are calculated on the basis of the model.
}
\end{figure}

We may expect intuitively that the most noticeable variation in the energy with the relative phase of the field components happens for $B_p \approx B_d$. Fig. \ref{E1(phi)} shows the predicted variations in the energies $E'(f_1)$ and $E''(f_1)$ at the fundamental frequency with $\varphi$ for $|\mathcal{B}(f_1)|/B_d = 1.14$ %(strip) 0.726 (disk)
and $|\mathcal{B}(3f_1)|/|\mathcal{B}(f_1)| = 1/18$ and 1/9. In this and following figures, all energies are normalized to the energy of the magnetic moment of the ideal diamagnetic sample in the applied field. The plotted experimental data of $E'(f_1)$ vs. $\varphi$, measured with $|\mathcal{B}(f_1)| = 200$ $\mu$T and $f_1=1.5625$ Hz at $T=88.76$ K, are in good agreement with the prediction. But also data for $|\mathcal{B}(3f_1)|/|\mathcal{B}(f_1)| = 1/6$ and 2/9, the values for which the model does not apply, show an identical character. While behavior of the variation in the experimental $E''(f_1)$ vs. $\varphi$ is in agreement with the prediction too, all experimental $E''(f_1)$ are higher - roughly by $0.16$ of $E''(f_1)$ in the pure sinusoidal field. This incongruity may be ascribed to the slightly higher experimental $\mathcal{M}_{1}''(f_1)$ compared with the predicted one, see Fig. \ref{m(B)}. This deviation is not systematic because it changes with different samples \cite{Janu2014}.

When the field component $\mathcal{B}(3f_1)$ is added in-phase, i.e. $\varphi = 0$, with $\mathcal{B}(f_1)$, the mean stored energy is increased while the energy dissipated per cycle of the fundamental component of the field is decreased compared with the energy in the pure sinusoidal field $\mathcal{B}(f_1)$. The effect rises with increasing $|\mathcal{B}(3f_1)|/|\mathcal{B}(f_1)|$ ratio. With increasing relative phase $\varphi$, the mean stored energy $E'(f_1)$ decreases while the energy $E''(f_1)$ dissipated per cycle of the field $\mathcal{B}(f_1)$ increases. The minimum in $E'(f_1)$ shifts from $\varphi \approx \pi$ to $\varphi \approx 3\pi/2$ in the process of increasing $|\mathcal{B}(3f_1)|/|\mathcal{B}(f_1)|$. The maximum in $E''(f_1)$ occurs for the field components in anti-phase, i.e. $\varphi = \pi$. As $\varphi$ approaches $2\pi$, both $E'(f_1)$ and $E''(f_1)$ return to its values for $\varphi = 0$.

\begin{figure}
\includegraphics[scale=0.45]{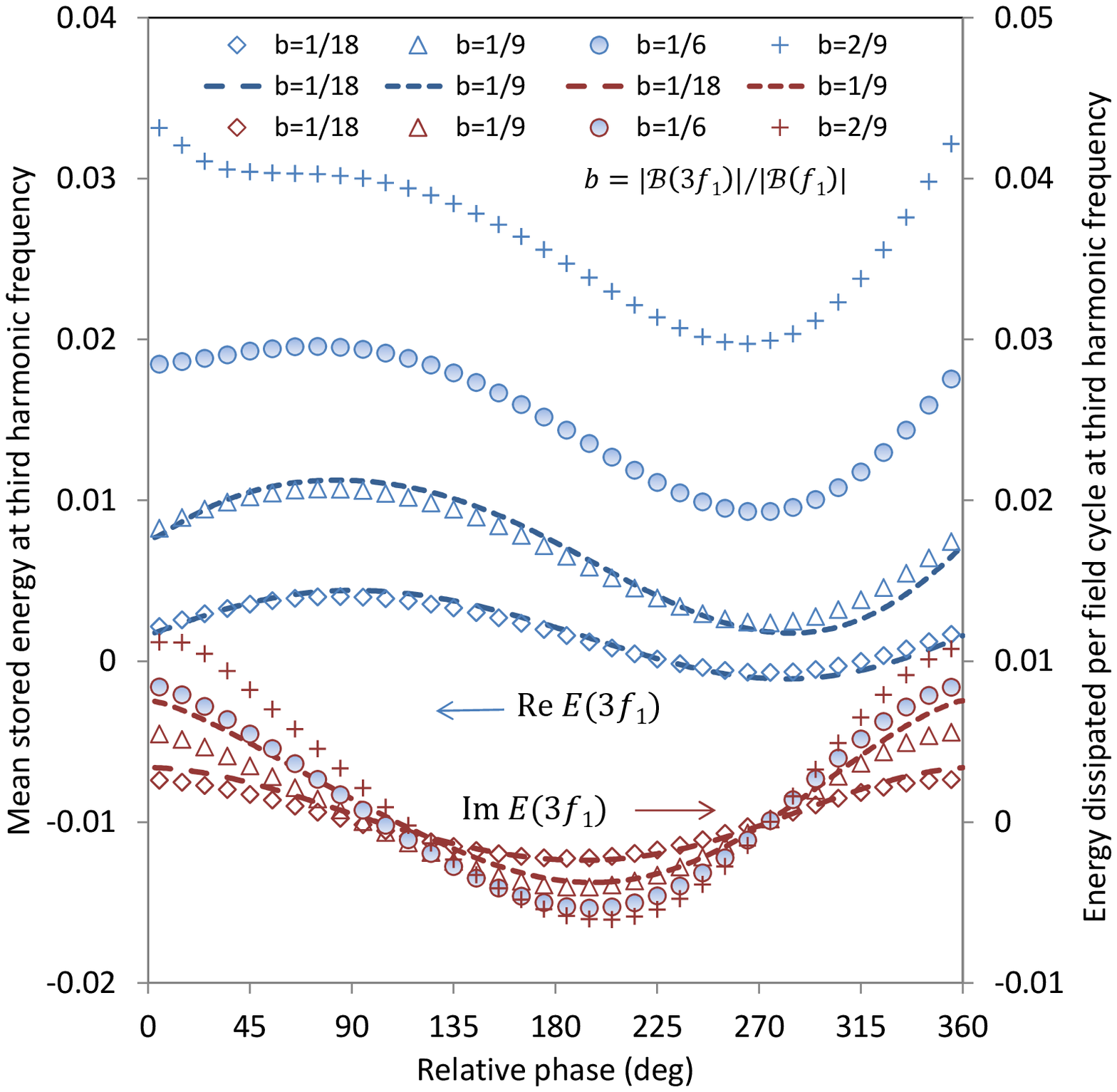}%{FigX1(TDivTc),X3(TDivTc)V3}% Here is how to import EPS art
\caption{\label{E3(phi)} The energy of the induced magnetic moment in the applied field at the third-harmonic-frequency vs. the relative phase $\varphi$ of the field components $\mathcal{B}(f_1)$ and $\mathcal{B}(3f_1)$ at fundamental frequency and third harmonic frequency, respectively. The symbols are for the experimental data. The curves are calculated on the basis of the model.
}
\end{figure}

The experimental and predicted variations in the energy at the third harmonic of the fundamental frequency, $E'(3f_1)$ and $E''(3f_1)$, with $\varphi$ are in good agreement, see Fig. \ref{E3(phi)}.

\begin{figure}
\includegraphics[scale=0.45]{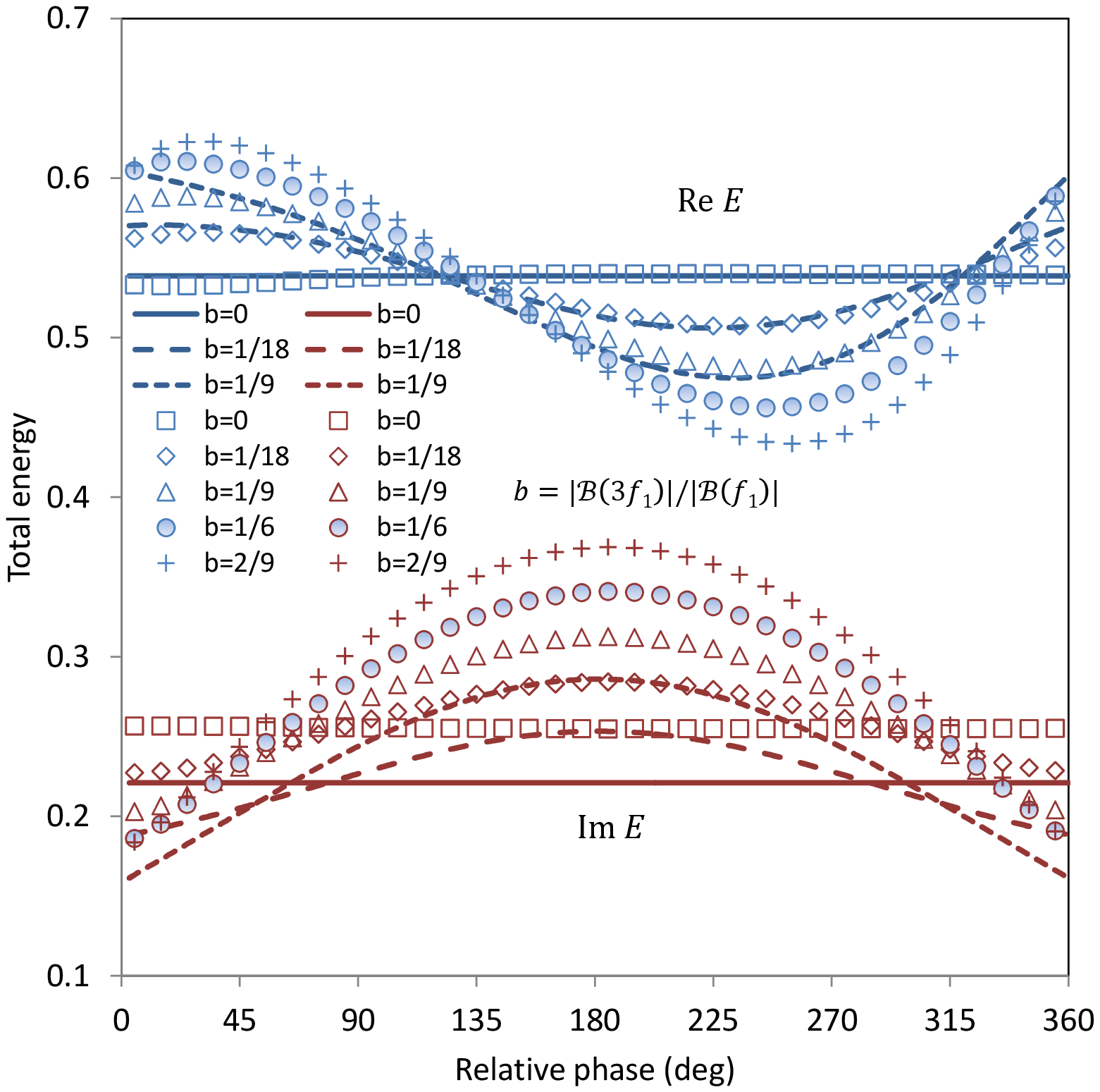}% Here is how to import EPS art
\caption{\label{TotalE(phi)} The total energy of the induced magnetic moment in the applied field vs. the relative phase $\varphi$ of the field components $\mathcal{B}(f_1)$ and $\mathcal{B}(3f_1)$ at fundamental frequency and third harmonic frequency, respectively. The symbols are for the experimental data. The curves are calculated on the basis of the model.
}
\end{figure}

Fig. \ref{TotalE(phi)} shows the variation in the total energy, $E'$ and $E''$, with relative phase $\varphi$ of the field components. As the $E(3f_1)$ is roughly an order of magnitude lower than $E(f_1)$ even for $|\mathcal{B}(3f_1)|/|\mathcal{B}(f_1)| = 2/9$, the chart is similar to that in Fig. \ref{E1(phi)}.

\begin{figure}
\includegraphics[scale=0.45]{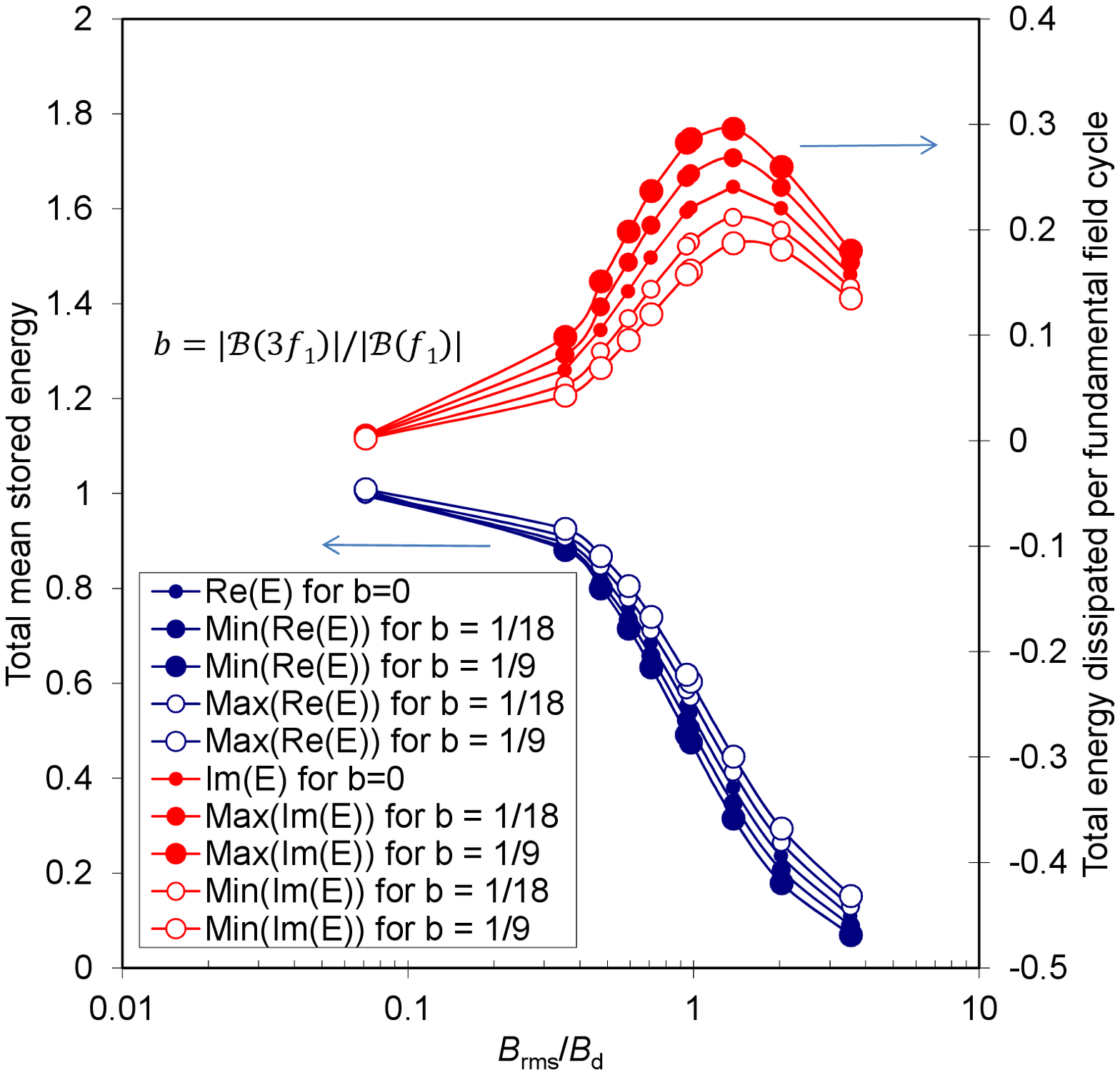}%{FigX1(TDivTc),X3(TDivTc)V3}% Here is how to import EPS art
\caption{\label{ExtremeValues} Normalized theoretical extreme values of the total mean stored energy and total energy dissipated per fundamental field cycle as a function of $B_{rms}/B_d$ calculated on the basis of the model.
}
\end{figure}

The predicted extreme values of the total stored energy and total energy dissipated per fundamental field cycle are shown in Fig. \ref{ExtremeValues}. The best effect on the variation in the energy with relative phase occurs for $B_{rms} \approx B_{d}$.

\section{Conclusion}

In conclusion, we have shown that the energy of a system with hysteretic nonlinearity in an applied periodic field, both the mean stored energy and energy dissipated per cycle of the fundamental component of the applied field varies with relative phase of the sinusoidal fields at harmonics of the fundamental frequency. A physical origin of hysteresis is well understood in the studied system. The complete analytical expressions known for the initial magnetization curves and hence for the full hysteresis loops offer an effective way to study this system. These expressions result from the critical-state-theory in type-II superconductors described by a single parameter, the critical depinning current density. The question is, for example, whether such a dissipative system and fluctuating field coupled to it may "self-organize". Fundamentals of linear-response theory, e.g. a fluctuation-dissipation theorem, are violated here.

\section*{Acknowledgements}
This work was partially supported by the Program of Czech Research Infrastructures (Project No. LM2011025).

{}

\end{document}